\documentclass{article}
\usepackage[utf8]{inputenc}
\usepackage[T1]{fontenc}
\usepackage{arxiv}
\usepackage{amsmath,amssymb}
\usepackage{graphicx}
\usepackage{booktabs}
\usepackage{array}
\usepackage[round,authoryear,sort&compress]{natbib}
\usepackage[hidelinks]{hyperref}
\usepackage{url}
\graphicspath{{./}{figures/}}

\title{Inunda: A GPU-Native, Differentiable Solver for High-Resolution Flood Inundation Modeling}
\author{%
Zhi Li\textsuperscript{1}\\
\textsuperscript{1}Civil, Environmental, and Architectural Engineering, University of Colorado Boulder, Boulder, CO, 80304\\
\textsuperscript{1}Institute of Arctic and Alpine Research, University of Colorado Boulder, Boulder, CO, 80304%
}

\begin{document}
\maketitle

\begin{abstract}
Predicting where floodwater goes and how deep it gets, at high resolution and across large domains, remains computationally expensive with conventional hydraulic solvers, while purely data-driven surrogates are fast but lack physical guarantees and generalize poorly beyond their training events. We present \textbf{Inunda}, a GPU-native flood inundation model that solves the two-dimensional shallow water equations entirely in PyTorch. Inunda uses a mass-conservative local-inertial scheme and runs multi-day events over millions of cells in minutes on a single GPU. Because every operator is autograd-compatible, the solver is \emph{differentiable by construction}: model parameters (surface roughness, channel bed elevation, and soil infiltration) can be estimated by gradient descent against gage observations through reverse-mode automatic differentiation of the full simulation. We demonstrate Inunda on three case studies. For a hindcast of Hurricane Harvey (2017) in Harris County, Texas, Inunda matches surveyed high-water marks to a mean absolute error of 0.67 m, competitive with or better than a suite of established flood models, and reaches a median gage water-level Nash--Sutcliffe efficiency of +0.72, more than double the +0.31 of the operational National Water Model v3.0. For the July 2025 Central Texas flash flood, Inunda is driven by an 18-member 1-km convection-allowing precipitation ensemble to produce probabilistic flood forecasts whose skill improves systematically as lead time to the crest shortens. For a post-fire flash-flood application in the Rio Ruidoso burn scar, differentiable calibration recovers the saturated hydraulic conductivity as a spatially explicit field at the model's own resolution and traces its multi-year post-fire recovery, which translates into a roughly threefold reduction in simulated flooded area as infiltration capacity returns. Inunda provides an end-to-end pipeline for real-event flood modeling that couples the accuracy of physics-based hydraulics with the calibration and coupling advantages of modern differentiable programming.
\end{abstract}
\keywords{flood inundation \and  shallow water equations \and  differentiable}
\section{Introduction}

Floods are among the most frequent and damaging natural hazards, and both their frequency and their consequences are rising as a warming climate intensifies extreme rainfall and as urban expansion places more people and assets in harm's way \citep{teng2017flood}. Climate projections indicate that river hydrographs are becoming flashier worldwide \citep{zhu2026flashiness} and that regions such as the conterminous United States are growing more prone to flash floods under high-end emissions \citep{li2022flashfloods}. Reducing flood impacts depends on knowing, at fine spatial resolution, \emph{where} water will go and \emph{how deep} it will get, information that underpins early warning, evacuation planning, infrastructure design, and risk pricing. Producing that spatially explicit picture, quickly enough to be useful for forecasting and at resolutions fine enough to resolve streets and channels, remains a central challenge for flood science.

The established route to such predictions is physics-based hydrodynamic modeling, in which the two-dimensional shallow water equations (SWEs) are solved over a digital elevation model to route water across the floodplain. Full-momentum solvers (e.g., HEC-RAS 2D, TELEMAC, Delft3D-FM) capture the relevant hydraulics faithfully, but their cost scales strongly with resolution, domain size, and event duration, so their computational demand grows non-linearly with model complexity \citep{teng2017flood}. To make large, high-resolution problems tractable, the community has widely adopted simplified formulations, most prominently the local-inertial (diffusive-inertial) scheme of LISFLOOD-FP, which neglects the advective momentum term while retaining local acceleration and gravity \citep{bates2010simple,deAlmeida2012improving}. This scheme underlies continental- and global-scale hazard products \citep{wing2017fathom}, reduced-physics coastal compound-flood solvers such as SFINCS \citep{leijnse2021sfincs}, and coupled hydrologic--hydraulic frameworks that pair a distributed rainfall--runoff model with a 2D inundation solver to map and predict flooding for real events, including catastrophic events such as Hurricane Harvey \citep{li2021crest,chen2021comprehensive}. A parallel line of work has ported hydrodynamic solvers to graphics processing units (GPUs), where the structured, data-parallel nature of a grid-based SWE update maps naturally onto massively parallel hardware \citep{kalyanapu2011assessment,xia2019hipims,moraleshernandez2021triton}. These GPU codes achieve large speedups, but they are typically written as bespoke CUDA or C++ kernels that stand apart from the machine-learning software ecosystem, and their parameters (such as surface roughness, infiltration, and channel geometry) are still calibrated by black-box search or manual tuning.

At the same time, data-driven and machine-learning (ML) approaches have advanced rapidly as an alternative to solving the governing equations directly. Deep neural networks have been trained as flood-inundation surrogates and emulators, using convolutional networks to predict inundation depth and extent \citep{kabir2020deep}, neural operators such as the Fourier neural operator for rapid inundation forecasting \citep{sun2023rapid}, recurrent (LSTM) networks for rainfall--runoff and streamflow \citep{kratzert2018rainfall}, and, at operational scale, ML systems that now deliver flood forecasts across large and ungaged regions \citep{nevo2022flood,nearing2024global}; recent reviews survey the breadth of these methods \citep{bentivoglio2022deep}. Once trained, such models are extremely fast. However, they carry well-documented limitations. Purely data-driven surrogates typically require large training corpora generated by the very hydrodynamic simulators they aim to replace; they generalize poorly when applied to locations or forcing regimes outside their training distribution; and they offer no intrinsic guarantee of physical constraints such as mass conservation.

These two threads are beginning to converge through \emph{differentiable programming}, in which a numerical simulator is written so that gradients of its outputs with respect to its inputs and parameters are available by automatic differentiation (AD), unifying physical models with machine learning and turning calibration and coupling into gradient-based problems rather than black-box search \citep{shen2023differentiable}. General-purpose AD frameworks have made whole PDE solvers differentiable and GPU-executable \citep{holl2020learning,bezgin2023jaxfluids}, and in hydrology differentiable formulations have reframed parameter estimation as learnable, scalable "parameter learning" that outperforms traditional calibration \citep{tsai2021calibration}. Yet, to date, this paradigm has not been brought to bear on a GPU-native, mass-conservative two-dimensional hydrodynamic flood solver capable of simulating real multi-day events at high resolution, leaving a gap between the physical fidelity of hydrodynamic models, the speed of GPU and ML approaches, and the calibration power of differentiable programming.

This paper introduces \textbf{Inunda}\footnote{Project website, documentation, and interactive demonstrations: \url{https://inunda.ai}.}, a flood inundation solver that closes that gap by combining physics-based hydraulics with the modern AI compute stack. Inunda solves the 2D shallow water equations with a mass-conservative local-inertial scheme \citep{bates2010simple} implemented \emph{entirely} in PyTorch tensor operations \citep{paszke2019pytorch}, with no custom CUDA kernels; it therefore runs GPU-native on the same hardware and software that powers deep learning, and simulates multi-day events over millions of cells in minutes on a single GPU. Because every operator is autograd-compatible, Inunda is \emph{differentiable by construction}: reverse-mode automatic differentiation through the solver's time loop enables gradient-based, pixel-wise estimation of Manning roughness, channel bed elevation, and CREST soil-infiltration parameters directly against gage observations. The model ships as an end-to-end pipeline, with built-in downloaders for terrain, rainfall, and validation data driven by a single configuration, and we validate it on real events against USGS streamgages and high-water marks and benchmark it against state-of-the-art flood models. In combining physical fidelity, GPU-native speed, and differentiability in one solver, Inunda offers both a fast forward model and a platform for gradient-based parameter estimation and for coupling with learned components.

The remainder of the paper is organized as follows. Section 2 describes the governing equations, the numerical scheme, the model forcing, and the differentiable calibration framework. Section 3 presents three case studies: a Hurricane Harvey (2017) hindcast benchmarked against other models and USGS observations, flood forecasting during the 2025 Central Texas floods, and post-fire flash-flood prediction with gradient-based parameter estimation. Section 4 discusses capabilities and limitations, and Section 5 concludes.

\section{Methods}

\subsection{Governing equations}

Inunda solves the two-dimensional shallow water equations in local-inertial (diffusive-inertial) form \citep{bates2010simple,deAlmeida2012improving}. The continuity equation is written in flux form,

\begin{equation}
\frac{\partial h}{\partial t} + \frac{\partial q_x}{\partial x}
+ \frac{\partial q_y}{\partial y} = R - I,
\end{equation}

where $h$ is water depth, $q_x, q_y$ are unit-width discharges (m\textsuperscript{2}/s), and $R$ and $I$ are the rainfall source and infiltration/loss sink. The momentum equation retains local acceleration and neglects advection,

\begin{equation}
\frac{\partial q}{\partial t}
= -\,g\,h\,\frac{\partial (z_b + h)}{\partial x}
  -\,\frac{g\,n^2\,q\,|q|}{h^{7/3}},
\end{equation}

with gravity $g$, bed elevation $z_b$, and Manning roughness $n$. Gravity is treated explicitly through the free-surface gradient $\nabla(z_b + h)$ so that terrain slope routes flow and uniform rainfall channelizes into topographic lows; Manning friction is treated semi-implicitly. A critical-flow cap (Froude $\le 1$) is applied on each face for stability.

\subsection{Numerical scheme}

The solver advances the state $(h, q)$ on a single uniform-resolution block covering the whole DEM, using a staggered arrangement in which depths $h$ are stored at cell centers and unit-width discharges $q$ on the cell faces between neighbours. Each time step is carried out in two sweeps. The flux sweep updates each face discharge from the current depth field by discretising the momentum equation, evaluating gravity explicitly from the free-surface gradient $\nabla(z_b + h)$ across the face and treating the Manning friction term semi-implicitly so that the friction response remains stable at large time steps \citep{bates2010simple,deAlmeida2012improving}. The depth sweep then updates each cell's depth from the net divergence of its face fluxes. Because the depth change of a cell is computed directly from the fluxes crossing its faces, water leaving one cell across a shared face is exactly the water entering its neighbour: the scheme is mass-conservative by construction, and total volume is conserved. All operators are expressed as pure PyTorch tensor shifts and elementwise operations, with no custom CUDA kernels, and the per-step update is fused with \texttt{torch.compile} so that the solver executes efficiently on the same GPU software stack used for deep learning \citep{paszke2019pytorch}.

\textbf{Stability controls.} Two mechanisms keep the explicit update stable. A critical-flow cap limits each face discharge so that the local Froude number does not exceed unity, preventing runaway acceleration on steep faces. In addition, the time step is chosen adaptively to satisfy the Courant--Friedrichs--Lewy (CFL) condition, $\Delta t = C\,\Delta x / \sqrt{g\,h_{\max}}$, where $\Delta x$ is the grid spacing, $h_{\max}$ the current maximum depth, and $C$ the Courant number; $\Delta t$ is capped at a configured maximum and clipped so that steps land exactly on the requested output times. For real events containing deep, incised channels, a Courant number of $C = 0.4$ maintains stability through the recession limb, where a larger value can destabilize the deepest channel cells.

\textbf{Steep terrain.} On steep or noisy topography, raw DEM cells can produce unphysically large free-surface gradients. Inunda conditions the DEM and clamps the local CFL time step where the terrain gradient is extreme, and can optionally represent buildings as elevated or blocked cells, so that flow is routed around structures rather than through them.

\textbf{Boundary conditions.} Real events use transmissive (open, zero-gradient) boundaries, implemented as a critical-flow free overfall that lets water leave the domain so the flood can recede toward its downstream outlet; closed no-flow (reflective) boundaries, in which water leaves only through losses, are also available.

\subsection{Forcing, losses, and inflow}

\textbf{Rainfall.} Rainfall enters the continuity equation as the source term $R$, supplied either as a spatially uniform constant rate over a specified duration or as time-stamped gridded heterogeneous quantitative precipitation estimates (QPE). For real events, native-resolution NOAA Multi-Radar Multi-Sensor (MRMS) radar QPE \citep{zhang2016mrms} is used, mapped onto the model grid through a precomputed index rather than being reprojected at every step. Because gravity is evaluated from the free-surface gradient, rain falling on the grid accumulates and channelizes into topographic lows on its own, without a prescribed drainage network.

\textbf{Infiltration and evaporation losses.} Losses are represented by the sink term $I$ in two ways. The simplest is a constant infiltration (and optional evaporation) rate subtracted from the ponded depth each step, capped so that depth never becomes negative. The physically richer option embeds the CREST (Coupled Routing and Excess Storage) model \citep{wang2011crest,li2023decadal} as an integrated land-surface model within the solver. CREST uses a variable-infiltration curve to partition incoming rainfall into surface runoff and soil storage on a per-cell basis, according to rainfall intensity and the evolving soil-moisture state that it carries as a prognostic variable. Crucially, CREST enables run-on infiltration: rather than treating infiltration as a one-time loss at the moment of rainfall, it re-infiltrates ponded surface water back into the soil column as overland flow routes across the domain, using the same mass-exact, water- and rate-limited kernel that governs the surface exchange. Runoff generated on one cell can therefore infiltrate on a downslope cell it flows onto, so the flood continues to lose water to the soil through the recession limb, a re-infiltration process shown to materially affect flood inundation mapping and prediction during extreme storms in the Texas Gulf Coast region \citep{li2022can}. CREST's soil parameters ($W_m$, $B$, $I_m$, $K_{sat}$) are spatially distributed fields, and, because the CREST computation is itself autograd-compatible, they can be estimated pixel-wise by the differentiable calibration described in Section 2.4.

\textbf{Distributed roughness.} Bed friction may be prescribed as a single uniform Manning coefficient $n$ or as a spatially distributed field. In the distributed case, each cell's roughness is assigned from its land-cover class using the USGS National Land Cover Database, mapped to engineering roughness values, so that a forested floodplain, a paved surface, and a channel each carry a physically appropriate friction. This spatial roughness is also one of the fields that can be calibrated by gradient descent.

\textbf{Fluvial inflow.} Inunda is forced by rain falling on the grid (pluvial forcing), so a river draining a large catchment that lies \emph{outside} the DEM contributes no water, and the domain-total volume is short by exactly that external inflow. To close the water balance on such domains, observed discharge is injected as additional channel inflow: each specified inflow location adds its time-varying volumetric flux to the lowest-lying (channel) cells nearby, spread over enough cells that the per-cell fill rate stays within the CFL limit, and the injected volume integrates back to the specified discharge exactly. Discharge may come from observed USGS streamgages or, where no gage exists, from modelled GEOGLOWS reach discharge, so the same mechanism serves gaged and ungaged basins.

\subsection{Differentiable calibration}

Because every operator is autograd-compatible, Inunda can be calibrated by reverse-mode automatic differentiation \emph{through the solver's time loop}. The objective minimized in this study (though the framework is not limited to this form) is

\begin{equation}
\mathcal{L}(\theta) =
\sum_{g}\sum_{t}
\frac{\big(\mathrm{rise}_{\mathrm{sim},g}(t;\theta)
           - \mathrm{rise}_{\mathrm{obs},g}(t)\big)^2}{N}
\;+\; \lambda\,\lVert \theta - \theta_{\mathrm{prior}} \rVert_1
\;+\; \lambda_{\mathrm{time}}\,\mathrm{timing}(\theta),
\end{equation}

where $\theta$ is one or more per-cell parameter fields: the Manning roughness $n$, the channel bed-elevation adjustment $dz$, and/or the CREST soil grids ($W_m$, $B$, $I_m$, $K_{sat}$). The first term is the data misfit between simulated and observed water-surface rise, pooled over the calibration gages and magnitude-weighted so that a near-dry gage cannot dominate; the second is an $L_1$ penalty that keeps each field close to its physical prior (the NLCD roughness map, the prior DEM, or the prior soil grids) and thereby regularizes the otherwise under-determined inverse problem; and the third is an optional timing penalty on the wave's arrival time or shape, which targets errors in \emph{when} a hydrograph rises rather than only its peak magnitude. The quantity compared to observations can be selected as event water-surface rise above a baseline, absolute water-surface elevation in a fixed vertical datum, or channel discharge integrated through a differentiable cross-section, the last being preferable where a fine channel cell's stage is confounded by local geometry.

\textbf{Tractable back-propagation.} The gradient $\partial\mathcal{L}/\partial\theta$ is obtained by reverse-mode automatic differentiation through the full unrolled simulation, which for an event of tens of thousands of adaptive steps requires care to remain numerically safe and within GPU memory. Inunda addresses this with three mechanisms. First, calibration runs a gradient-safe twin of the solver in which the argument of every square root that can reach zero on a dry face (the critical-flow cap) is floored to a small positive value, preventing the $0\cdot\infty$ products that would otherwise poison the backward pass; the parameters are the only leaf tensors and the forward engine is otherwise unchanged. Second, truncated back-propagation-through-time splits the event into chunks that carry the solver state forward detached, so the autograd graph spans only one chunk at a time. Third, within a chunk, runs of steps are wrapped in gradient checkpointing, recomputing the forward pass during back-propagation rather than storing every intermediate, which trades computation for a square-root reduction in memory. Together these hold peak memory to roughly 5--6 GB at 30 m resolution. Parameters are updated with the Adam optimizer using per-field learning rates, non-finite gradients are scrubbed, the per-field gradient norm is clipped, and each field is clamped to its physical range at every step; the fit retains the best iterate rather than the last. Finally, a parameter update can be confined to a binary mask, so that a field is re-tuned only where it is expected to have changed: for example, re-estimating infiltration and roughness only inside a wildfire burn scar while holding both at their prior values elsewhere. Because reverse-mode AD through the solver is memory-intensive, calibration is performed at 30 m resolution and the resulting parameter fields are then applied at the finer deployment resolution.

\subsection{Urban drainage coupling}

A rain-on-grid model routes all water over the surface, but a real city drains a large fraction of its runoff underground through the storm-sewer network before it reaches a street or channel; neglecting the sewers systematically over-predicts urban pluvial flooding for events below roughly the 10--50-year design storm. Inunda can therefore couple an urban storm-drain network that exchanges water with the 2D surface, in either of two modes, both restricted to the local-inertial solver and both leaving a case that omits them bit-for-bit unchanged.

The first is a capacity-limited inlet abstraction: ponded depth on network cells is drained at a prescribed inlet-capacity rate (rate- and water-limited by the same mass-exact kernel used for CREST run-on infiltration), banked into a pipe reservoir, and optionally released back at an outfall through a linear reservoir with a lag. Being closed-form in depth $h$, this mode is differentiable, so the inlet-capacity field can in principle be calibrated as another leaf parameter.

The second mode is a full bidirectional 1D/2D coupling to the EPA Storm Water Management Model (SWMM), in which the 2D surface and a real SWMM5 pipe network exchange water \emph{interactively} at every manhole. SWMM is driven in-process, and once per coupling interval Inunda gathers the surface water depth at each junction cell, computes a signed exchange discharge for that node according to the local hydraulic regime (surcharge from a pressurized pipe back onto the surface, or weir- and orifice-controlled inflow from the surface into the pipe), injects that discharge as the node's lateral inflow, and advances SWMM by one interval. The regime switching and its exchange coefficients follow established coupled urban-drainage practice, and numerical stability is maintained with sign-flip damping, under-relaxation against the previous exchange, and a hard limiter that prevents a node from intercepting more water than the surface cell holds. Because SWMM is an external black box, the exchange flows enter the differentiable surface update in a detached manner (as with the rainfall forcing), so gradients with respect to the surface parameters still propagate. Where municipal storm-drain GIS is unavailable, a complete SWMM network can be synthesized for the domain from open data. This coupling lets Inunda represent urban drainage scenarios in which the interaction between surface flooding and the sub-surface pipe network governs where and how deeply water ponds. A known limitation is tailwater feedback: a synthetic network's outfalls discharge freely, so during extreme events the pipes rarely surcharge as they would when the receiving water body is high and backs the system up into the streets (Section 4).

\begin{figure}[htbp]
\centering
\includegraphics[width=\textwidth]{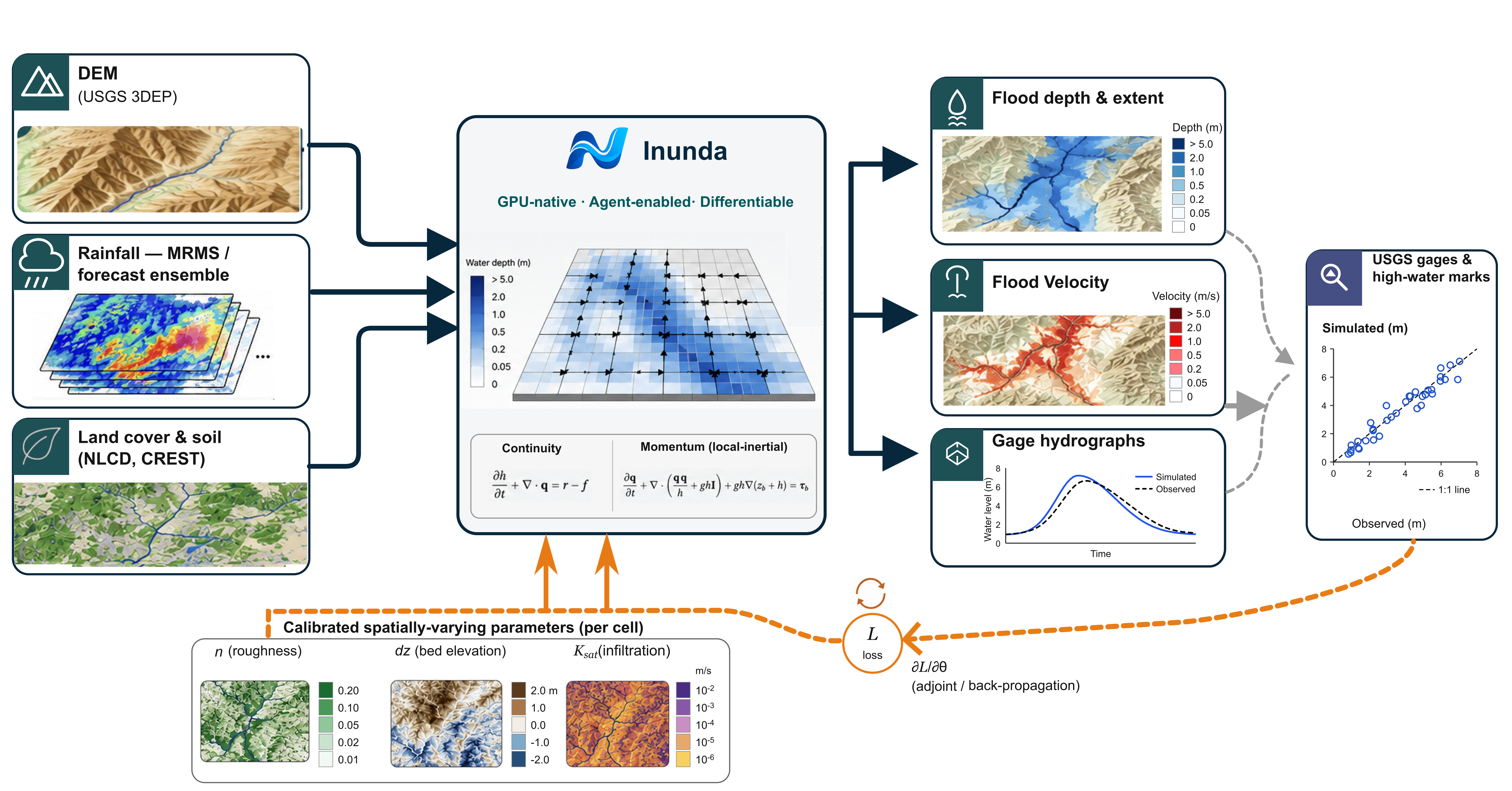}
\caption{Schematic of the Inunda pipeline. Gridded inputs (terrain from the USGS 3DEP DEM, rainfall from MRMS radar QPE or a forecast ensemble, and land cover and soil from NLCD and CREST parameters) force the GPU-native, mass-conservative local-inertial 2D shallow-water solver (governing continuity and momentum equations shown), which advances water depth and discharge on a structured grid and produces flood depth and extent, flow velocity, and gage hydrographs as CF-compliant output. Simulations are validated against USGS gages and surveyed high-water marks. The dashed loop indicates the differentiable-calibration path: a loss $\mathcal{L}$ between simulated and observed quantities is back-propagated through the solver ($\partial\mathcal{L}/\partial\theta$) to estimate spatially varying per-cell parameters: Manning roughness $n$, bed-elevation adjustment $dz$, and saturated hydraulic conductivity $K_{sat}$.}
\label{fig:1}
\end{figure}

\section{Case Studies}

\subsection{Flood inundation simulation during Hurricane Harvey}

Hurricane Harvey (August 2017) produced catastrophic multi-day rainfall over Harris County, Texas (basin rainfall on the order of 660 mm), making it a demanding test of a high-resolution flood solver and a well-observed event for which several published flood-model results exist. We simulate the event over the Harris County domain at 10 m resolution on a USGS NED digital elevation model, forced by NOAA MRMS radar QPE \citep{zhang2016mrms}. The domain is treated as an enclosed catchment with no external fluvial inflow, so all flood water derives from rain falling on the grid. To establish realistic antecedent conditions, we warm up the model by beginning the simulation three days before the event. We validate against two independent observation sets: continuous USGS streamgage records and surveyed high-water marks (HWMs).

\textbf{Spatial flood extent and depth.} Figure 2 shows the simulated maximum-depth field alongside three reference maps for the same event: a Fathom/LISFLOOD-FP simulation, the coupled hydrologic--hydraulic CREST-iMAP model \citep{li2021crest} (here applied to the Spring Creek watershed only, because CREST-iMAP solves the full-form shallow water equations and is therefore too computationally expensive to run over the entire domain at 10 m resolution), and the FEMA regulatory flood map. Inunda reproduces the county-wide inundation pattern (the main-stem bayous, the western reservoir pools, and the eastern San Jacinto floodplain) at a level of spatial detail comparable to Fathom. Panel (e) maps the signed difference between simulated and observed HWM depths at 488 marks, showing that errors are close to zero across much of the domain with no strong spatial bias.

\textbf{Gage hydrograph skill.} Simulated water levels are compared to USGS gage observations at 40 gages and scored by Nash--Sutcliffe efficiency (NSE). Because Inunda outputs water-surface stage and the NOAA National Water Model (NWM) v3.0 \citep{nwm2016} outputs channel discharge, each model is scored against its native observed variable: Inunda stage against USGS gage height (parameter 00065), and NWM discharge against streamflow (parameter 00060). Inunda achieves a median stage NSE of \textbf{+0.72}, more than double the NWM's median discharge NSE of \textbf{+0.31} over the same gages (Figure 3b). The spatial distribution of skill (Figure 4) shows Inunda positive at nearly all gages across the county, whereas the NWM is more spatially mixed with a substantial fraction of negative-NSE gages in the urban core.

\textbf{Benchmark against other flood models.} On the 813 USGS high-water marks in the domain, Inunda's peak-depth mean absolute error (MAE) is \textbf{0.67 m} and approximately unbiased, the lowest of the models shown in Figure 3a and second only to one published result (Huang et al. 2021, 0.65 m, not mapped here) among the eight models compiled for this event. It sits below the FEMA observation-reference map (0.70 m), the coupled CREST-iMAP (0.81 m), the published SFINCS Harvey skill (0.83 m \citep{sebastian2021harvey}), Fathom/LISFLOOD-FP (1.05 m \citep{wing2017fathom,bates2010simple}), and Delft3D-FM (1.34 m). In a direct head-to-head on common riverine HWMs, Inunda's error is below that of the downloaded Fathom grid, whose own error matches its published value and thereby validates the comparison. We acknowledge that this comparison is not fully controlled: we did not obtain and re-run each model under an identical configuration, but instead compiled the best available simulated maximum-flood- depth product for each model from publicly released datasets and evaluated all of them against the same high-water marks.

\begin{table}[htbp]
\centering
\caption{Peak-depth mean absolute error (MAE) against USGS high-water marks in the Harris County domain (813 marks; Inunda \textasciitilde{}unbiased). Values recomputed in this study are marked accordingly; others are published skill for the same event.}
\label{tab:1}
\begin{tabular}{lcl}
\toprule
Model & HWM depth MAE (m) & Source \\
\midrule
\textbf{Inunda} & \textbf{0.67} & this study \\
FEMA (obs-reference) & 0.70 & this study \\
CREST-iMAP & 0.81 & this study \\
SFINCS (Sebastian 2021) & 0.83 & published \\
Fathom / LISFLOOD-FP & 1.05 & this study \\
Delft3D-FM (Lee 2024) & 1.34 & published \\
\bottomrule
\end{tabular}
\end{table}

\begin{figure}[htbp]
\centering
\includegraphics[width=0.9\textwidth]{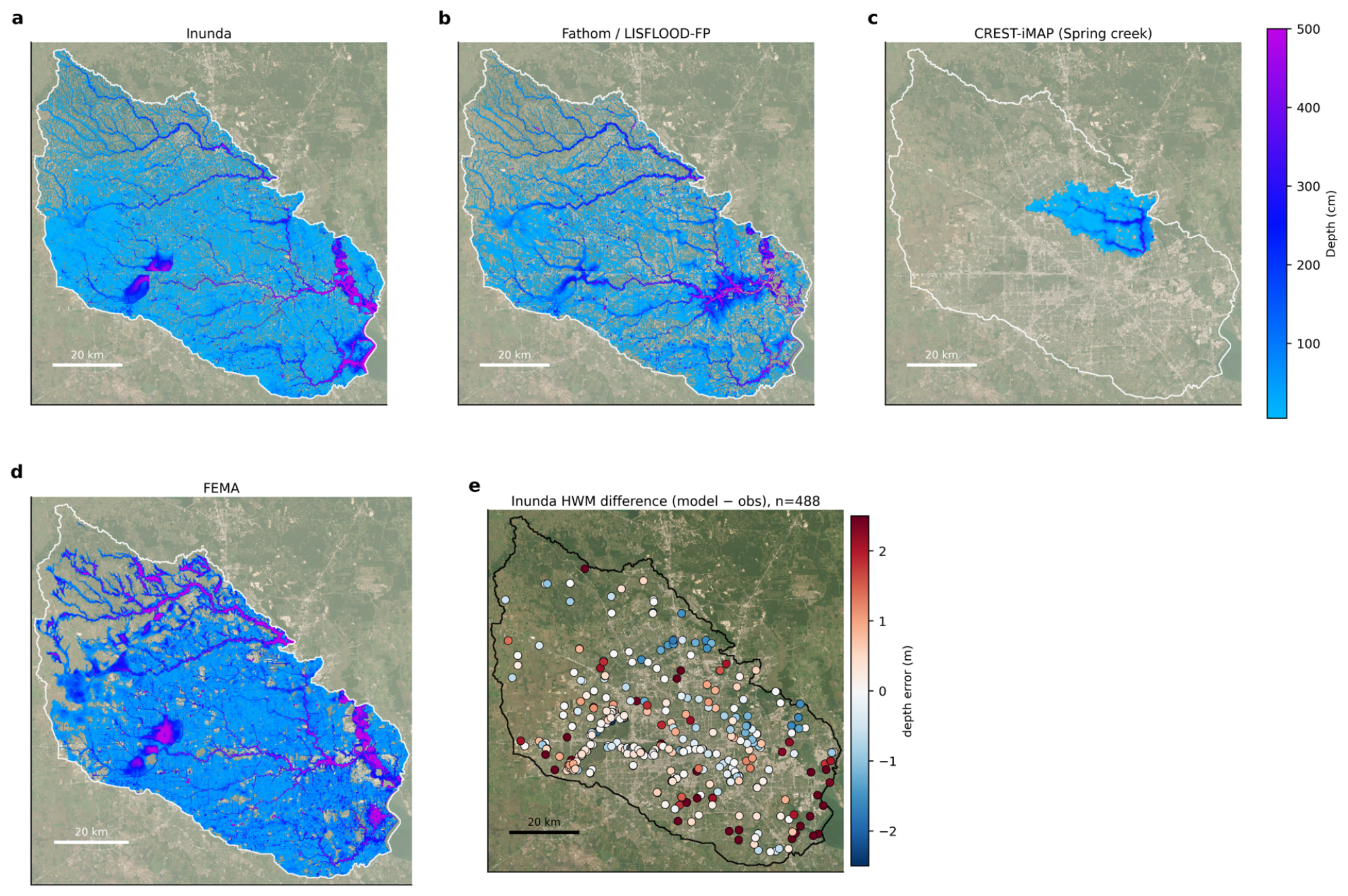}
\caption{Hurricane Harvey (2017) maximum flood depth over the Harris County domain. (a) Inunda; (b) Fathom/LISFLOOD-FP; (c) CREST-iMAP (Spring Creek watershed); (d) FEMA regulatory flood map; color gives water depth (cm). (e) Signed difference between Inunda and observed high-water-mark depths (model $-$ observation, m) at 488 marks.}
\label{fig:2}
\end{figure}

\begin{figure}[htbp]
\centering
\includegraphics[width=0.9\textwidth]{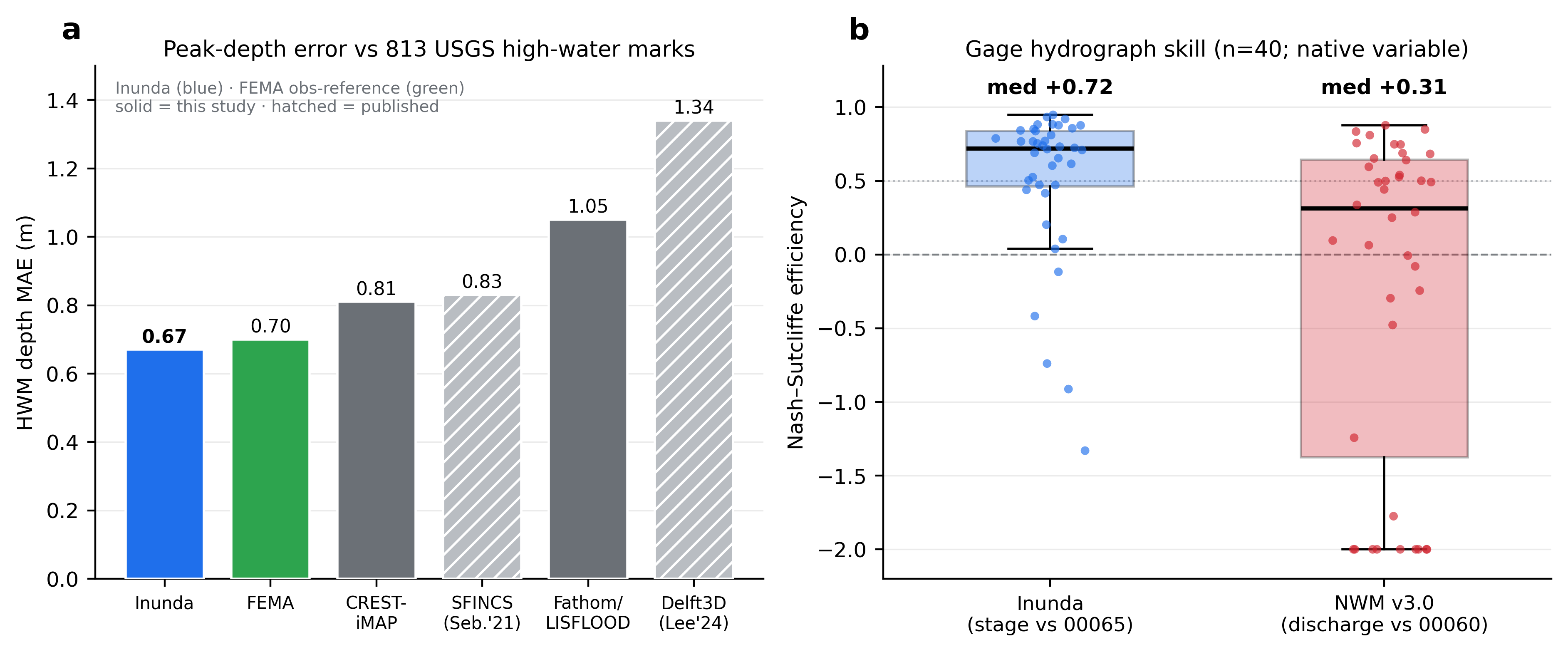}
\caption{Inunda skill on Hurricane Harvey. (a) Peak-depth MAE against 813 USGS high-water marks for Inunda (blue), the FEMA observation-reference (green), and four benchmark flood models (gray; solid = recomputed in this study, hatched = published). (b) Gage hydrograph skill (Nash--Sutcliffe efficiency) at 40 gages, each model in its native variable: Inunda stage (vs USGS parameter 00065, median +0.72) and NWM v3.0 discharge (vs parameter 00060, median +0.31).}
\label{fig:3}
\end{figure}

\begin{figure}[htbp]
\centering
\includegraphics[width=0.9\textwidth]{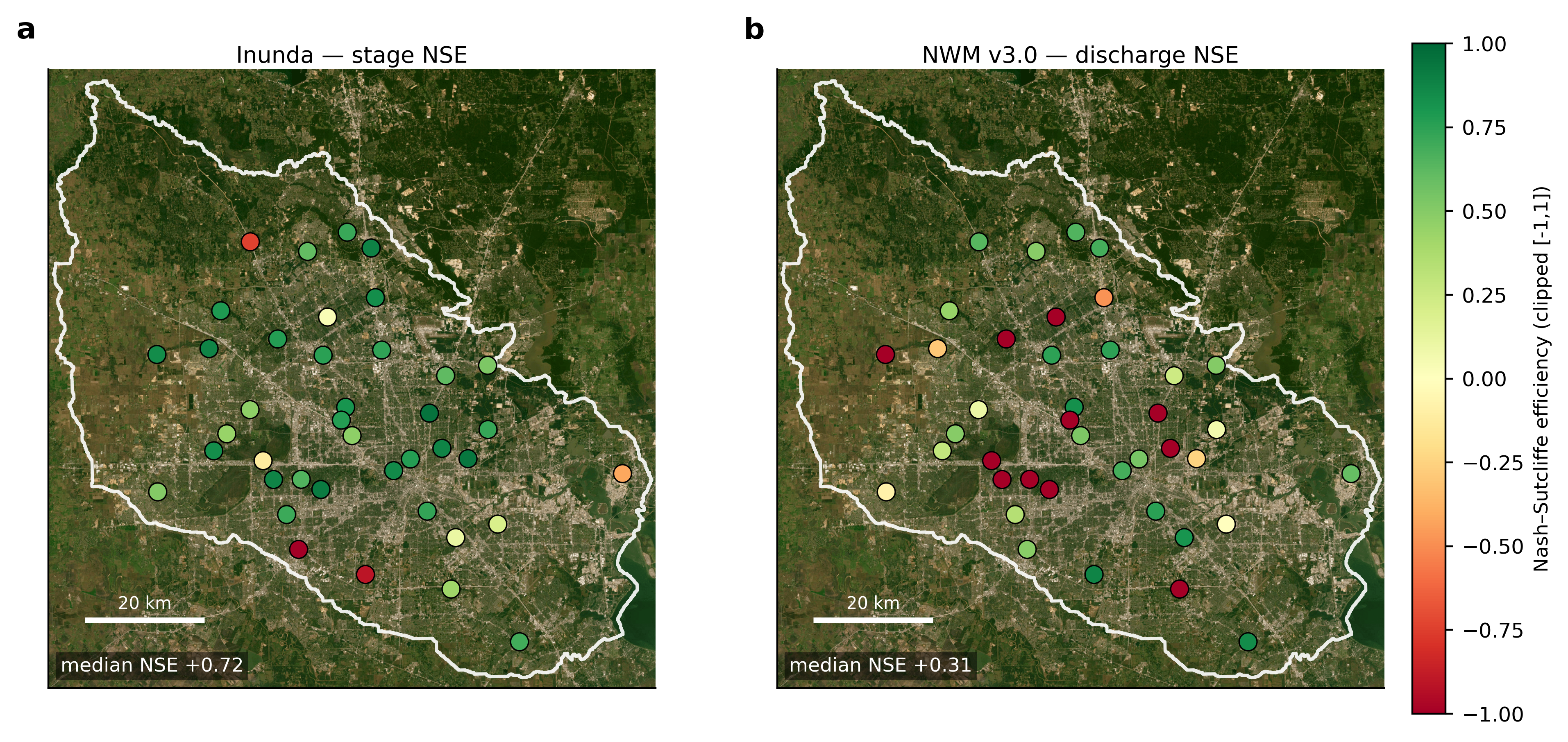}
\caption{Spatial distribution of gage hydrograph skill across the Harris County domain. (a) Inunda stage NSE (median +0.72); (b) NWM v3.0 discharge NSE (median +0.31). Marker color gives NSE clipped to [$-$1, 1].}
\label{fig:4}
\end{figure}

\textbf{Computational performance across GPU hardware.} Because Inunda is written entirely in standard PyTorch tensor operations with no custom CUDA kernels, the same code runs unmodified on any CUDA-capable GPU and inherits each new generation's hardware improvements without porting effort. To isolate hardware throughput, we benchmarked one representative sub-basin of the Harvey domain, Brays Bayou (USGS hydrologic unit 120401040402; about 119 km\textsuperscript{2}, 1.2 million cells at 10 m resolution), over the full six-day Harvey window (25--31 August 2017) on five data-center and workstation GPUs spanning three NVIDIA architectures (Table 2). Every GPU produced an identical integration (949,079 adaptive time steps at a mean step of 0.546 s), so the differences in run time reflect pure hardware throughput rather than any change in the numerics. On the reference A100-SXM4-80GB the sub-basin solves in 254 s (313 s wall-clock including the cached terrain fetch and gage and map validation), advancing at roughly 3,700 solver steps per second. The newer Hopper (H100, H200) and Blackwell (B200, RTX PRO 6000 workstation card) parts are all faster, with the two Blackwell cards fastest at about 11,500 steps per second, a 3.1-fold speedup over the A100. The ranking does not, however, follow chip generation strictly: the H100 (2.0$\times$ over the A100) outpaces the newer H200 (1.4$\times$) on this workload. This behaviour is consistent with the memory-bandwidth-bound nature of the stencil update, in which achievable memory throughput on the specific workload, rather than peak arithmetic rate, sets performance; the two Blackwell cards, for instance, deliver nearly identical solve times despite very different peak floating-point rates. That a multi-day simulation of this size completes in minutes, and continues to accelerate with each hardware generation at no code cost, is what makes running an entire forecast ensemble (Section 3.2) feasible within the lead time of a flash flood.

\begin{table}[htbp]
\centering
\caption{Computational performance benchmark on the Brays Bayou sub-basin (hydrologic unit 120401040402; \textasciitilde{}119 km\textsuperscript{2}, 1.2 M cells at 10 m) over the full six-day Harvey window, run on Modal (PyTorch 2.8.0+cu128, driver 580.95). All GPUs executed the identical integration (949,079 adaptive steps, mean step 0.546 s). "Solve" is the time in the numerical solver; "wall" is the total run time including the cached terrain fetch and validation; "steps/s" is the mean adaptive time steps per second; speedup is relative to the A100 solve time.}
\label{tab:2}
\begin{tabular}{lcccc}
\toprule
GPU & Solve (s) & Wall (s) & Steps/s & Speedup vs A100 \\
\midrule
B200 & 82.5 & 108.0 & 11,501 & 3.08$\times$ \\
RTX PRO 6000 (Blackwell) & 83.1 & 110.0 & 11,415 & 3.06$\times$ \\
H100 80GB HBM3 & 124.9 & 160.1 & 7,597 & 2.03$\times$ \\
H200 & 179.3 & 232.1 & 5,293 & 1.42$\times$ \\
A100-SXM4-80GB & 253.9 & 312.7 & 3,738 & 1.00$\times$ \\
\bottomrule
\end{tabular}
\end{table}

\subsection{Flood forecasting in the 2025 Central Texas floods}

The 4 July 2025 Central Texas (Texas Hill Country) flash-flood disaster provides a forecasting test case that exercises Inunda in a fundamentally different mode from the Harvey hindcast: rather than being forced by observed radar rainfall after the fact, the model is driven by \emph{forecast} precipitation and must anticipate the flood before it occurs. As the forcing, we use the 1-km convection-allowing ensemble precipitation forecasts from the NOAA Warn-on-Forecast System (WoFS), whose coupled hydrometeorological forecasts for this event are described by \citet{gourley2026wofs}. Coupled hydrologic--hydraulic forecasting of extreme events has previously been demonstrated with high-resolution quantitative precipitation forecasts and deep-learning precipitation nowcasts as forcing \citep{chen2022flood}; here the GPU-native runtime of Inunda allows the full 2D hydrodynamic forward model to be run for every ensemble member inside the short window a flash flood allows. We simulate the Kerrville domain on the Guadalupe River at 30 m resolution and verify against the USGS gage at Hunt, Texas (08165500). One consequence of the 30 m grid is that it under-resolves the confined river channel, which limits the absolute stage the model can attain (below).

\textbf{Probabilistic flood forecasting.} Each of the 18 WoFS ensemble members is used to force an independent Inunda simulation, and the resulting inundation fields are combined into an ensemble flood-probability map, defined at each cell as the fraction of members simulating a depth greater than 0.1 m (Figure 5). Running the system from a sequence of nine forecast initializations at 30-minute intervals from 06:00 to 10:00 UTC (corresponding to lead times of 5.8 down to 1.8 hours ahead of the simulated flood peak) shows how the forecast sharpens as the event approaches. As the initialization time moves closer to the event, the high-confidence footprint, defined as the area exceeding 50\% ensemble probability A(P$\ge$0.5), contracts from roughly 95 km\textsuperscript{2} at the longest lead time to about 36 km\textsuperscript{2} at the shortest, as the members converge on a progressively more localized flood extent.

\begin{figure}[htbp]
\centering
\includegraphics[width=0.9\textwidth]{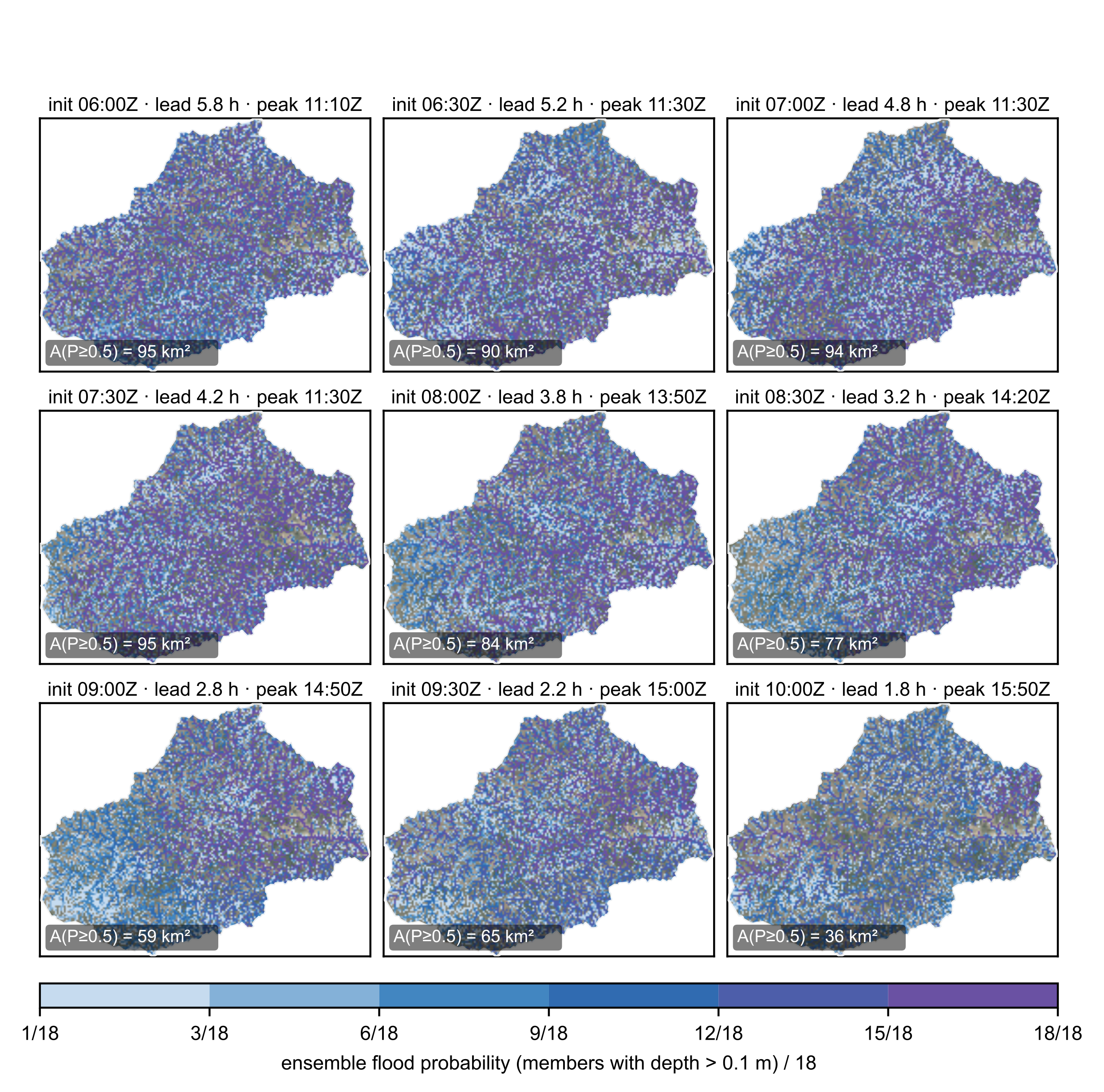}
\caption{Ensemble flood-probability maps for the July 2025 Central Texas flood, from Inunda driven by the 18-member WoFS 1-km precipitation ensemble \citep{gourley2026wofs}. The nine panels show successive forecast initializations at 30-minute intervals from 06:00 to 10:00 UTC (decreasing lead time to the flood peak, 5.8 h to 1.8 h); color gives the ensemble flood probability (fraction of the 18 members with depth > 0.1 m). The inset in each panel reports the area exceeding 50\% probability, A(P$\ge$0.5).}
\label{fig:5}
\end{figure}

\textbf{Lead-time dependence of forecast skill.} Figure 6 verifies the forecasts against the observed water-surface elevation at the USGS gage at Hunt, Texas (08165500), where the flood crested at 499.5 m NAVD88. As a reference, an analysis run forced by observed MRMS radar QPE reaches a peak of approximately 496 m, about 3.5 m below the observed crest, so this analysis, rather than the observed stage, represents the practical skill ceiling set by the model, its resolution, and the forcing. Part of this gap is that the 30 m terrain does not resolve the subgrid heterogeneity of the confined channel, so the station's local topography (and hence the peak stage attainable there) is under-resolved. Relative to that ceiling, the WoFS-forced ensemble forecasts improve monotonically as the lead time to the crest shrinks: forecasts initialized roughly 5--6 hours ahead substantially under-predict the rise and its timing, whereas forecasts initialized about 2 hours ahead recover nearly the full MRMS-forced analysis peak (Figure 6b). Each forecast is warm-started from the MRMS-forced antecedent state at its initialization time, so the ensemble spread reflects the divergence of the forecast rainfall alone rather than differences in the initial condition.

\begin{figure}[htbp]
\centering
\includegraphics[width=0.9\textwidth]{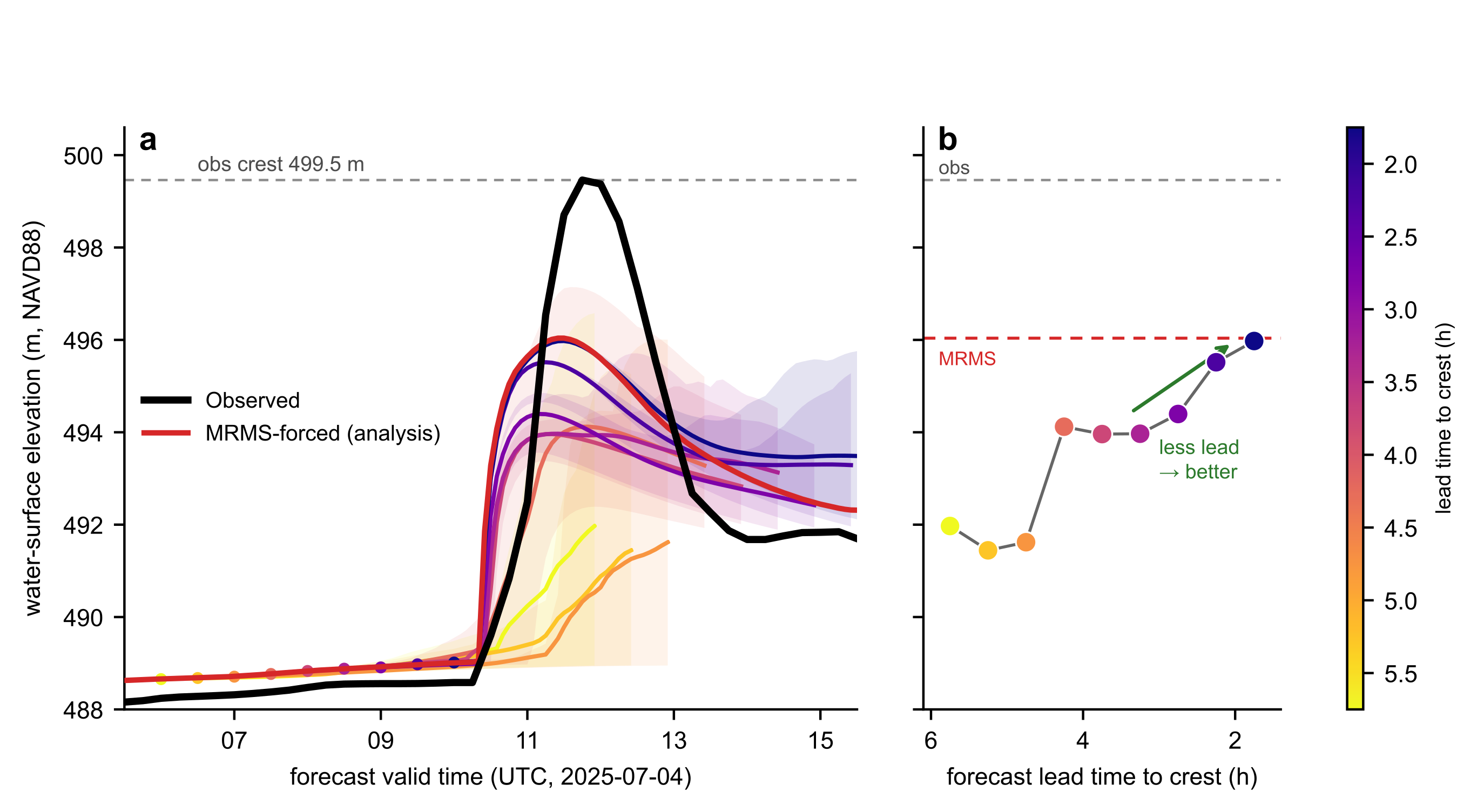}
\caption{Lead-time dependence of the 2025 Central Texas flood forecast at the USGS gage at Hunt, Texas (08165500), where the flood crested at 499.5 m NAVD88. (a) Water-surface elevation through time: observed (black), the MRMS-forced analysis (red), and the WoFS-forced ensemble members colored by their lead time to the crest; shading shows the ensemble spread. (b) Peak water-surface elevation as a function of forecast lead time to the crest, showing convergence toward the MRMS-forced reference (red dashed) as lead time decreases; the observed crest (gray dashed) sits above the model ceiling.}
\label{fig:6}
\end{figure}

\subsection{Post-fire flash-flood prediction with parameter estimation}

Wildfire burn scars collapse soil infiltration capacity, producing flashier, earlier runoff and flash floods whose behavior conventional, static parameters cannot represent. The controlling parameter is the saturated hydraulic conductivity $K_{sat}$, which drops sharply where fire consumes the litter layer and induces soil water repellency, and then recovers over subsequent years as vegetation and soil structure re-establish. This case study exploits the differentiability of the framework to treat $K_{sat}$ not as a fixed input but as a field to be \emph{learned} from observed flood response, and to do so at the model's own simulation resolution rather than the kilometre-scale of conventional soil-parameter products.

Because the loss can be back-propagated to a per-cell $K_{sat}$ field, calibration resolves the burn-scar infiltration structure at the same fine resolution as the hydrodynamic simulation, replacing a spatially uniform or coarsely mapped prior with a spatially explicit estimate. Repeating the estimation for successive post-fire years recovers a \emph{recovery trajectory}: the year-by-year re-establishment of infiltration capacity inside the burn scar. Confining the update to a burn-scar mask (Section 2.4) ensures that only the fire-affected soil is re-estimated while the surrounding, unburned landscape is held at its prior.

We apply this to the Rio Ruidoso burn scar in south-central New Mexico, which reached its maximum extent following the 2019 fire. The burn-scar mask is derived from the differenced Normalized Burn Ratio (dNBR), and the model (and hence the recovered $K_{sat}$ field) is run at 30 m resolution, so that infiltration capacity is estimated at the same fine scale as the flood simulation rather than being read from a kilometre-scale soil product. For each post-fire year, the per-cell $K_{sat}$ field is calibrated by back-propagating a loss against observed gage discharge, so that the recovered infiltration capacity is the field that best reproduces that year's observed streamflow response.

Figure 7 shows the recovered $K_{sat}$ field for each year from the year of maximum scar through the most recent year, together with a static (no-recovery) reference. Within the burn scar, the mean recovered infiltration capacity rises from 3.5 mm h\textsuperscript{-}\textsuperscript{1} in the year of maximum scar (2019), identical to the static no-recovery reference, to 5.8, 6.0, 7.0, and 7.1 mm h\textsuperscript{-}\textsuperscript{1} over the following four years (2020--2023), before a partial setback to 6.7 mm h\textsuperscript{-}\textsuperscript{1} in 2024 and a further rise to 8.7 mm h\textsuperscript{-}\textsuperscript{1} in 2025. The 2024 reversal interrupts an otherwise monotonic recovery and is consistent with a second fire in the watershed that year, which would have renewed soil water repellency and depressed infiltration capacity before recovery resumed. The maps make the spatial character of this recovery visible: infiltration capacity returns first and most strongly in parts of the scar, while other areas remain suppressed years after the fire, a heterogeneity that a single basin-average recovery curve would hide.
\begin{figure}[htbp]
\centering
\includegraphics[width=0.9\textwidth]{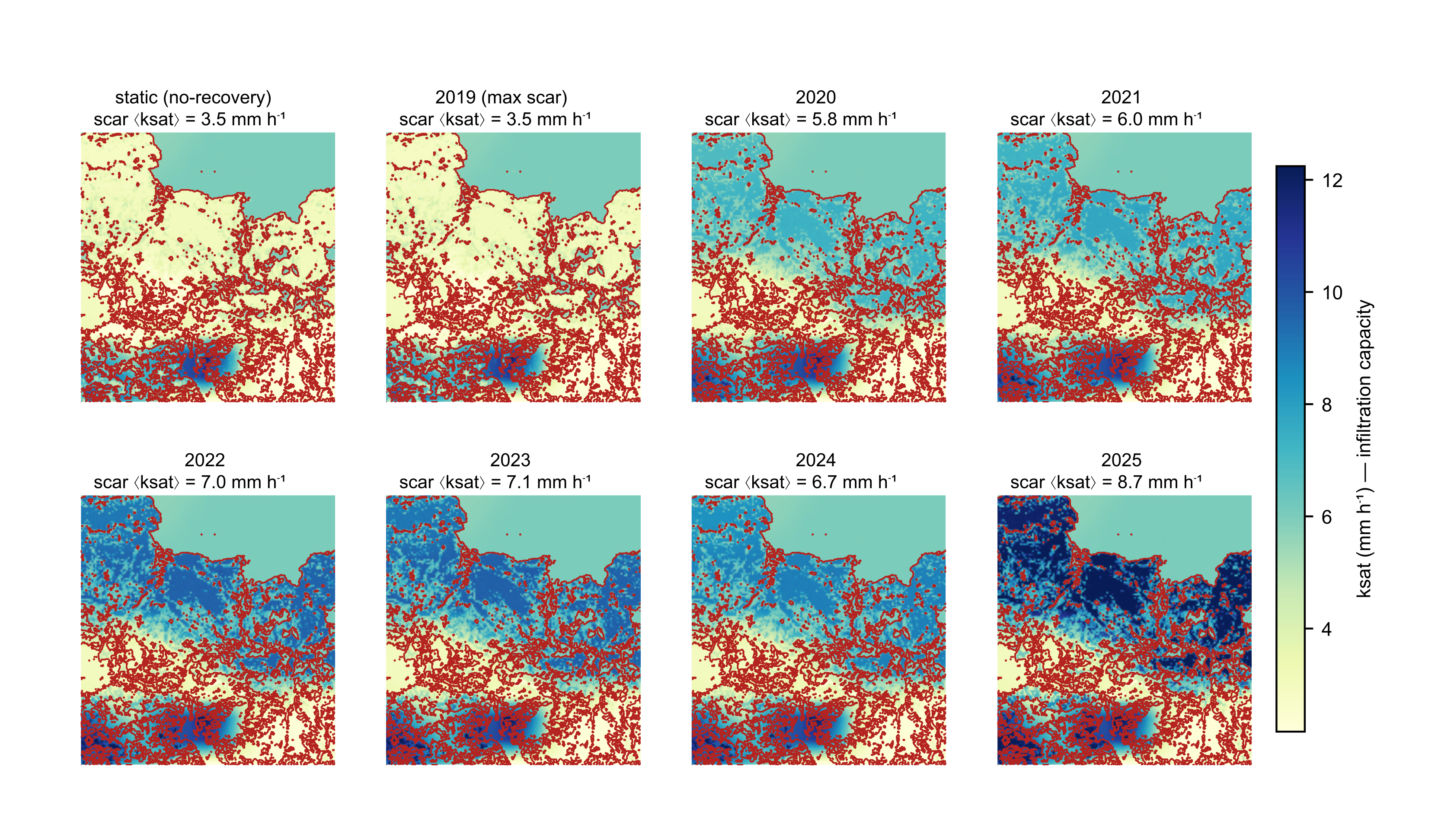}
\caption{Differentiable recovery of post-fire saturated hydraulic conductivity $K_{sat}$ inside a wildfire burn scar. Each panel maps the recovered per-cell $K_{sat}$ (color; mm h\textsuperscript{-}\textsuperscript{1}, infiltration capacity) for a given year, with the burn-scar boundary outlined; panels run from a static no-recovery reference and the year of maximum scar (2019) through 2025. Titles report the scar-mean $K_{sat}$, $\langle K_{sat}\rangle$, which increases from 3.5 mm h\textsuperscript{-}\textsuperscript{1} at maximum scar to 8.7 mm h\textsuperscript{-}\textsuperscript{1} by 2025, tracing the post-fire recovery trajectory at the model's simulation resolution.}
\label{fig:7}
\end{figure}

\textbf{Flood-response consequence.} Because the recovered $K_{sat}$ fields feed back into the forward model, the recovery trajectory has a direct hydrological signature: driving Inunda with each year's parameters under a common design storm translates the infiltration recovery into a decline in simulated flooding (Figure 8). Immediately post-fire, the depressed infiltration produces a flooded area (depth > 0.1 m) of 1.93 km\textsuperscript{2}, identical to the static no-recovery reference; as infiltration recovers, this contracts to 0.76 km\textsuperscript{2} by 2020 and to 0.63 km\textsuperscript{2} by 2025, roughly a threefold reduction in flooded area attributable to soil recovery alone. The trajectory is not strictly monotonic: flooded area ticks back up to 0.70 km\textsuperscript{2} in 2024, mirroring the same-year setback in recovered $K_{sat}$ and underscoring that the flood response tracks the learned infiltration field year-by-year. This closes the loop of the case study: a differentiably learned, spatially resolved soil parameter is carried back through the physics to yield a physically consistent, quantitative post-fire flood-hazard trajectory.

\begin{figure}[htbp]
\centering
\includegraphics[width=0.9\textwidth]{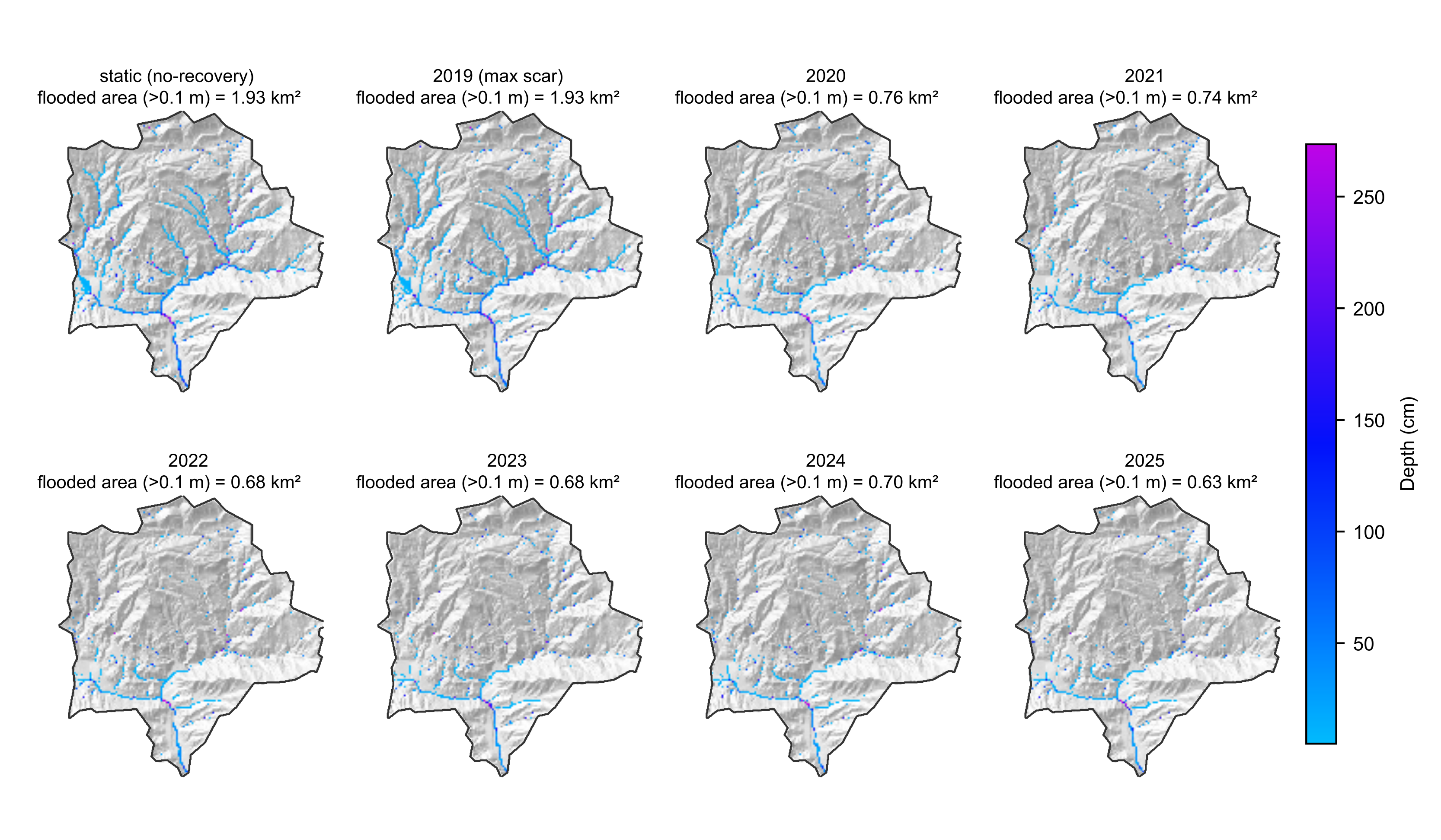}
\caption{Simulated flood response to post-fire soil recovery in the Rio Ruidoso burn scar. Each panel shows the Inunda maximum-depth field (color; cm) under a common design storm, forced by the recovered $K_{sat}$ field for the corresponding year (Figure 7), over a static no-recovery reference and 2019--2025. Titles report the flooded area (depth > 0.1 m), which contracts from 1.93 km\textsuperscript{2} immediately post-fire to 0.63 km\textsuperscript{2} by 2025 as infiltration capacity recovers, with a slight 2024 reversal that mirrors the recovered-$K_{sat}$ setback.}
\label{fig:8}
\end{figure}

\section{Discussion}

Taken together, the three case studies illustrate two properties that follow directly from Inunda's design. The first is that expressing a mass-conservative hydrodynamic solver in the same tensor framework used for deep learning makes it fast enough to change how it is used. The Harvey hindcast shows that a GPU-native local-inertial solver attains high-water-mark and gage skill that is competitive with established flood models on a demanding real event; the Central Texas case shows that the same solver is fast enough to run an entire precipitation ensemble within the lead time of a flash flood, turning a deterministic hindcast tool into a probabilistic forecasting one. The second property is that differentiability turns model calibration and model coupling from outer-loop search problems into gradient problems. The post-fire case makes this concrete: rather than tuning a handful of lumped parameters, the framework recovers a spatially explicit soil-infiltration field at the model's own resolution and, by repeating the estimation year by year, extracts a physically interpretable recovery trajectory that feeds back through the physics into a quantitative flood-hazard signal. Because the solver, its land-surface model, and its calibration all live in one autograd-compatible pipeline with open, CF-compliant inputs and outputs, the same machinery serves hindcasting, forecasting, parameter learning, and coupling with learned components.

These strengths come with limitations that bound where the model should be trusted. Inunda solves the local-inertial approximation to the shallow water equations, which neglects advective momentum; it is therefore best suited to the gradually varied floodplain and channel flows that dominate large inundation events, and is not a substitute for a full-momentum solver in strongly supercritical or rapidly varied flow. Its skill is bounded by the resolution of the terrain and the fidelity of the forcing: on the Central Texas event, the 30 m DEM under-resolves the confined Guadalupe River channel and caps the peak stage the model can attain, and in general reservoir and lake margins are only as accurate as the available bathymetry, which conventional topographic DEMs do not capture below the water surface. The rain-on-grid formulation resolves pluvial flooding directly but requires external discharge to be injected where a river drains a catchment outside the domain, so the domain boundary must be chosen with the contributing area in mind. The urban-drainage coupling captures the first-order effect of the storm-sewer network, but a synthetic network whose outfalls discharge freely omits the tailwater feedback that occurs when a high receiving water body backs the system up into the streets, so it can over-drain the surface during the most extreme events. Finally, because the model is mass-conservative and adds no subgrid wetting, it is conservative about flood \emph{extent}: it is accurate where it predicts water but tends to under-spread relative to observed inundation, an effect most visible for lighter events and at the wet-boundary fringe.

Several of these limitations point to natural extensions. The differentiable core makes it straightforward to couple learned components (for example a neural correction to the friction or infiltration fields, or a learned closure for subgrid wetting) and to train them end-to-end against observations through the solver rather than in isolation. The same autograd interface that enables parameter learning also enables tighter, physically consistent coupling with external models, of which the interactive SWMM coupling is a first example. Because the computational cost is dominated by memory bandwidth rather than arithmetic, distributing the domain across multiple GPUs is the direct route to larger and finer applications. More broadly, a differentiable flood solver is a natural substrate for the emerging generation of geospatial foundation models and learned parameterizations, providing the physical constraints and the gradients through which such components can be trained on real flood observations.

\section{Conclusion}

Inunda is a GPU-native, differentiable solver for the two-dimensional shallow water equations, written entirely in standard tensor operations so that a mass-conservative hydrodynamic flood model runs on the same hardware and software stack as modern deep learning. On a Hurricane Harvey hindcast it reproduces surveyed high-water-mark depths to a mean absolute error of 0.67 m and attains a median gage water-level Nash--Sutcliffe efficiency of +0.72, competitive with or better than established flood models and well above the operational National Water Model on the same gages. Its speed allows an entire convection-allowing precipitation ensemble to be propagated through the full hydrodynamics within the lead time of the 2025 Central Texas flash flood, yielding probabilistic forecasts that sharpen as the event approaches.

The model's defining feature is that it is differentiable by construction. This turns calibration into a gradient problem: in the Rio Ruidoso post-fire case, back-propagation through the full simulation recovers the saturated hydraulic conductivity as a spatially explicit field at the model's own resolution and traces its multi-year recovery, which propagates back through the physics into a threefold reduction in simulated flood extent as infiltration returns. Delivered as an open, end-to-end pipeline with standard geospatial inputs and CF-compliant outputs, Inunda unifies physics-based accuracy with the calibration and coupling capabilities of differentiable programming, and provides a physical, trainable substrate for the next generation of hybrid physics--machine-learning flood models.

\section*{Open Research / Data Availability}

All input data used in this study are publicly available. Digital elevation data are from the USGS 3D Elevation Program (3DEP; \url{https://www.usgs.gov/3d-elevation-program}), and land-cover data are from the USGS National Land Cover Database (NLCD; \url{https://www.usgs.gov/centers/eros/science/national-land-cover-database}). Precipitation forcing uses NOAA Multi-Radar Multi-Sensor (MRMS) quantitative precipitation estimates, and streamflow and high-water-mark observations are from the U.S. Geological Survey. The simulated maximum-flood-depth products for the benchmark models compared in Section 3.1 are available on HydroShare (\url{https://www.hydroshare.org/resource/fae24734d6fc47be8bf0b54d6a175d86/}), and the FEMA maximum-flood-depth reference map is available on HydroShare (\url{https://www.hydroshare.org/resource/165e2c3e335d40949dbf501c97827837/}). The Inunda model, documentation, and interactive demonstrations are available at \url{https://inunda.ai}.

\section*{Acknowledgments}

We thank the U.S. Geological Survey, NOAA, and FEMA for making these datasets freely and openly accessible, without which this work would not have been possible. A portion of the computing was performed under a small allocation from the NSF National Center for Atmospheric Research (NSF NCAR).

\bibliographystyle{agu}
\bibliography{references}

\end{document}